# Enhanced Biaxial Compressive Strain Tuning of 2D semiconductors via Hot Dry Transfer on Polymer Substrates

Alvaro Cortes-Flores[1,2,#], Eudomar Henríquez-Guerra[1,2,#,*], Lisa Almonte[1,2], Hao Li[3], Andres Castellanos-Gomez[4], M. Reyes Calvo[1,2,5,*]

[1] BCMaterials, Basque Center for Materials, Applications and Nanostructures, 48940 Leioa, Spain

[2] Instituto Universitario de Materiales de Alicante (IUMA), Universidad de Alicante, Alicante 03690, Spain

[3] Nanoscale Physics and Devices Laboratory, The Institute of Physics, Chinese Academy of Sciences, P.O. Box 603, 100190 Beijing, China

[4] 2D Foundry Research Group, Instituto de Ciencia de Materiales de Madrid (ICMM-CSIC), Madrid E-28049, Spain

[5] IKERBASQUE, Basque Foundation for Science, 48009 Bilbao, Spain

# equal contribution

* Email: eudomar.henriquez@bcmaterials.net, reyes.calvo@bcmaterials.net

## Abstract

Strain engineering is an effective tool for tailoring the properties of two-dimensional (2D) materials, especially for tuning quantum phenomena. Among the limited methods available for strain engineering under cryogenic conditions, thermal mismatch with polymeric substrates provides a simple and affordable strategy to induce biaxial compressive strain upon cooling. In this work, we demonstrate the transfer of unprecedentedly large levels of uniform biaxial compressive strain to single-layer $WS_2$ by employing a pre-straining approach prior to cryogenic cooling. Using a hot-dry-transfer method, single-layer $WS_2$ samples were deposited onto thermally expanded polymeric substrates at 100 °C. As the substrate cools to room temperature, it contracts, inducing biaxial compressive strain (up to ~0.5%) in the $WS_2$ layer. This pre-strain results in a measurable blueshift in excitonic energies compared to samples transferred at room temperature, which serve as control (not pre-strained) samples. Subsequent cooling of the pre-strained samples from room temperature down to 5 K leads to a remarkable total blueshift of ~200 meV in the exciton energies of single-layer $WS_2$. This energy shift surpasses previously reported values, indicating superior levels of biaxial compressive strain induced by the accumulated substrate contraction of ~1.7%. Moreover, our findings reveal a pronounced temperature dependence in strain transfer efficiency, with gauge factors approaching theoretical limits for ideal strain transfer at 5 K. We attribute this enhanced efficiency to the increased Young's modulus of the polymeric substrate at cryogenic temperatures.

# INTRODUCTION

Strain engineering is a powerful approach for tuning the physical properties of two-dimensional (2D) materials[1,2]. While the effects of tensile strain have been widely studied, compressive strain remains comparatively less explored, particularly under cryogenic conditions. Yet, biaxial compressive strain has been predicted to significantly impact the quantum properties of 2D materials, such as excitonic phenomena, and low-temperature magnetic and superconducting phase transitions[1,3–5]. However, experimental validations of this impact remain limited, largely due to the lack of suitable methods capable of applying large, uniform compressive strain at low temperatures. One promising approach leverages the large thermal deformation of polymeric substrates[6–10] and the mismatch in thermal expansion coefficients between the polymer substrate and the 2D material deposited on top. In this approach, tensile or compressive strain is induced by heating or cooling the entire 2D-material/polymer system, respectively[6–8]. Recent work has demonstrated that this strategy can efficiently transfer substantial biaxial compressive strain to 2D materials by cooling down to cryogenic temperatures[10–12]. Polymers with high thermal expansion coefficients are thus desirable for inducing significant strain. Moreover, the efficiency of strain transfer from the polymer to the 2D material is strongly influenced by the substrate's mechanical properties, particularly its Young's modulus[13]. A higher Young's modulus of the substrate – relative to that of the 2D material – has been shown to result in more efficient strain transfer[7,13]. While Young´s modulus is known to change with temperature, the impact of this temperature dependence on strain transfer remains largely unexplored.

A promising but still largely unexplored approach to induce compressive strain involves transferring 2D materials onto flexible substrates that have been pre-expanded, for example by heating to moderately elevated temperatures. Upon cooling to room temperature, the thermal contraction of the substrate induces biaxial compressive strain in the 2D sample. This method was introduced by Kim et al.[14], who transferred few-layer black phosphorus onto a polyethylene terephthalate glycol modified (PETG) substrate at 95 °C, successfully achieving biaxial compressive strain upon cooling to room temperature. This approach allowed them to subsequently apply uniaxial tensile strain by bending the flexible substrate, effectively shifting the accessible strain range from compression to tension in a single experimental run at room temperature. Beyond this application, pre-straining samples by this method holds great potential for expanding the range of accessible compressive strain, particularly through further cooling to cryogenic temperatures, where additional strain can be induced by thermal contraction of the substrate. Despite its promise, the use of pre-straining methodologies to increase accumulated strain remains largely unexplored.

Two-dimensional semiconductors are highly sensitive to external perturbations due to their large surface-to-volume ratio, which allows their electronic properties to be readily tuned[15–18]. Within 2D semiconductors, transition metal dichalcogenides (TMDCs) exhibit exceptional mechanical strength[19–22], and enhanced Coulomb interactions lead to the formation of highly bound excitons, endowing this family of materials with robust optical properties[23–28]. These characteristics make single-layer TMDCs excellent candidates for strain-tuning their optical response[29–33]. In fact, strain has been shown to significantly modify the electronic band structure of TMDC semiconductors: compressive strain increases the bandgap energy, while tensile strain reduces it[1,34]. Various methods have been developed to study biaxial strain effects in single-layer semiconductors, including substrate bending[35–42], nanoindentation techniques[43–45], pressurizing single-layer membranes[46,47], and piezoelectric actuators[48–50]. Among these, the use of piezoelectric actuators is the only approach that can homogeneously and reversibly apply compressive strain across a wide temperature range. However, this technique is limited to small strain levels (~0.3%) [50]. In contrast, thermal mismatch with polymeric substrates has been shown to induce large amounts of uniform biaxial strain in single-layer TMDCs, both tensile by heating and compressive by cooling[7,8,10]. Among TMDCs, single-layer $WS_2$ has demonstrated the highest tunability of optical absorption under biaxial compressive strain, with excitonic shifts reaching ~130 meV per percent of induced strain[10]. This makes 1L-$WS_2$ an ideal platform for exploring advanced strain methodologies aimed at further enhancing its optical properties.

In this work, we achieved unprecedentedly high levels of uniform biaxial compressive strain in single-layer $WS_2$ by combining a hot-dry transfer at 100 °C with subsequent cooldown to cryogenic temperatures. First, 1L-$WS_2$ flakes were deposited onto thermally expanded polycarbonate substrates at 100 ºC. Upon cooling down to room temperature, the substrate contracted by ~0.5%, inducing biaxial compressive strain in the flakes, reflected in a ~30 meV blueshift in the excitonic peaks compared to control samples transferred at room temperature. Further cooling of these pre-strained samples from room temperature to 5 K results in an additional blueshift of the excitonic features—up to 200 meV—~50 meV larger than in the not pre-strained controls. This finding indicates an exceptionally high level of biaxial compressive strain induced by the accumulated substrate contraction (~1.7%), Interestingly, the gauge factor increased with decreasing temperature, reaching values as high as 148 meV/%, close to theoretical predictions (144–151 meV/%), suggesting enhanced strain transfer efficiency, which correlates with the rise in the Young's modulus of polycarbonate at cryogenic temperatures.

## RESULTS AND DISCUSSION

**Biaxial Compressive Strain at room temperature**

Single-layer $WS_2$ (1L-$WS_2$) samples were deposited on polycarbonate (PC) substrates using two distinct dry transfer methodologies (see the **Materials and Methods** section). On the one hand, a conventional dry transfer method[51] was employed to deposit the 1L-$WS_2$ flake onto a PC substrate at room temperature (RT, 23ºC or 296 K). On the other hand, a hot-dry-transfer method was implemented to deposit the 1L-$WS_2$ flake onto a previously heated PC substrate at 100 °C (373 K). This process was conducted at a maximum temperature of 100 °C to remain safely below the glass transition temperature of polycarbonate (139-147 ºC)[52,53].

A schematic of the hot transfer method is presented in **Figure 1a**. The single-layer $WS_2$ sample was first identified by optical inspection and micro-reflectance spectroscopy on a poly(dimethylsiloxane) (PDMS) substrate (see **Methods**). Then, the flake-PDMS system was brought into contact with the pre-heated PC substrate at 100 °C (373 K). After allowing the system to thermalize for 15 minutes, the PDMS was slowly peeled off, completing the transfer of the flake onto the PC substrate. The resulting flake–PC system was subsequently allowed to cool down to room temperature at a controlled rate of approximately 1 K/min to minimize the risk of slippage. We characterized the thermal expansion of polycarbonate between RT and 100 °C (see **Supporting Information, Section S1**), following the method reported in previous works[7,8,10]. Notably, the PC substrate expands by ~0.5% when heated up from RT to 100 °C and contracts by the same amount when cooled back to room temperature. This value is consistent with the thermal expansion coefficient of PC (~7 × 10$^{-5}$ K$^{-1}$) and the temperature interval. Therefore, upon cooling from 100 °C to RT, we expect the 1L-WS$_2$ flake to experience up to ~0.5% biaxial compressive strain (assuming perfect strain transfer), due to the mismatch in thermal expansion coefficients between the PC substrate and the 2D material (~5 × 10$^{-6}$ K$^{-1}$ and ~ 7 × 10$^{-5}$ K$^{-1}$, respectively)[8,54].

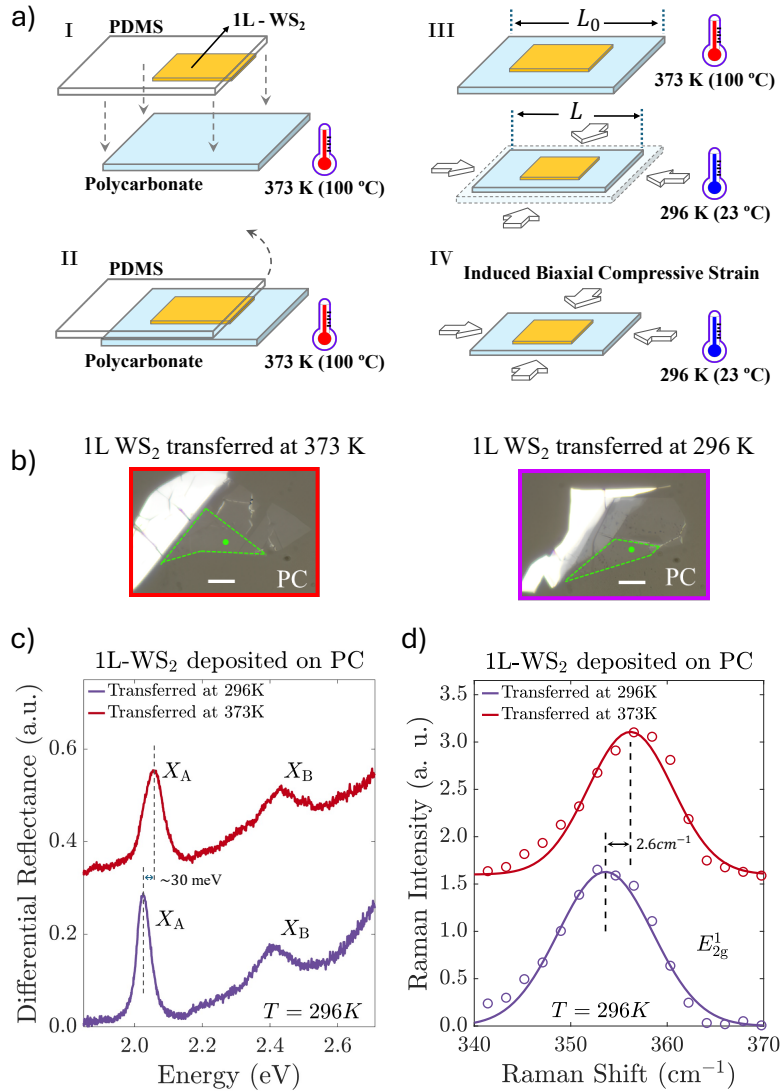

**Figure 1**. (a) Schematic diagram illustrating the steps involved in the deterministic hot transfer of a single-layer WS$_2$ flake onto a preheated polycarbonate (PC) substrate at 373 K (100ºC). (b) Optical images of the transferred flakes, with the single-layer region outlined by a green dashed line and green dots marking the ~3 μm reflectance measurement spots. White scale bars represent 10 μm. (c) Differential reflectance (DDR) and (d) Raman spectra measured at 296 K for a single-layer WS$_2$ flake on a PC substrate, transferred either via a deterministic dry transfer method at 296 K (23ºC, purple), or by the hot-transfer method at 373 K (100 ºC, red) shown in panel (a). Spectra are vertically offset for clarity.

To determine the effect of the induced strain, differential micro-reflectance (DDR) spectra (see **Methods**) were measured at room temperature (23ºC, 296 K) for both the RT-transferred and the hot-transferred 1L-WS$_2$ samples (see **Figure 1b,c**). This technique has been widely used to investigate excitonic phenomena and to quantify strain effects in the optical properties of single-layer TMDCs [7,10,35,36,39,55]. In the visible range, DDR spectra of 1L-WS$_2$ present two prominent resonances corresponding to the A and B excitons, labelled $X_A$ and $X_B$, respectively. Comparison between the spectra of RT- and hot-transferred samples (**Figure 1c**) reveals an energy blue shift of ~30 meV in the excitonic resonances for the sample transferred at 100 °C relative to the control transferred at room temperature. To further investigate the origin of this shift, Raman spectroscopy was performed at room temperature on both samples, revealing a shift of ~2.6 cm$^{-1}$ to higher wavenumbers in the E$^1_{2g}$ mode for the hot-transferred sample (**Figure 1d**), while the A$_{1g}$ mode remained mainly unaffected (see **Supporting Information, Figure S2**).

Since both DDR spectra—corresponding to the 1L-WS$_2$ samples transferred at 23 °C and 100 °C—were acquired at the same temperature (23 °C, 296 K), we attribute the observed difference in exciton energies to biaxial compressive strain transferred from the substrate to the monolayer during the cooling step following the hot-transfer process. This interpretation is further supported by the Raman measurements, also performed at room temperature, which revealed a shift to higher wavenumbers in the E$^1_{2g}$ mode for the RT-transferred sample, consistent with ~0.5% biaxial compressive strain, as the in-plain E$^1_{2g}$ mode is expected to be more sensitive to biaxial lattice compression than the out-of-plane A$_{1g}$ mode[46]. Furthermore, similar strain-induced shifts in exciton energies and Raman modes have been consistently observed in other 1L-WS$_2$ samples transferred onto PC substrates using the hot-transfer technique (see **Supporting Information, Tables S1** and **S2**), supporting the reproducibility of these effects. For simplicity, we henceforth refer to the sample transferred at 100 ºC (373 K) as pre-strained and the sample transferred at 23 ºC (296K) as not pre-strained.

To test the reversibility of strain-induced exciton modulation, we investigated the temperature evolution of the differential reflectance spectra in the range 23 -100 ºC (296-373 K) for both the pre-strained and not pre-strained samples, monitoring the energy shifts of their excitonic resonances (**Figure 2a-e**, see **Materials and Methods** section). To isolate strain effects from the thermal contributions to exciton energy shifts, we also fabricated a control 1L-WS$_2$ sample on a SiO$_2$/Si substrate—which exhibits a negligible thermal expansion coefficient compared to PC[10,56]—and characterized it via differential reflectance spectroscopy over the same temperature range (see **Supporting Information, Figure S3**). As the temperature increases from 23 to 100 ºC, the $X_A$ and $X_B$ excitonic resonances in the DDR spectrum of 1L-WS$_2$ on SiO$_2$/Si exhibit an almost linear redshift of ~30 meV, with a slope of ~-0.4 meV/K. This redshift is consistent with the expected bandgap reduction due to the thermal increase of lattice vibrations and electron–phonon interactions[57,58]. Given the moderate difference in thermal expansion coefficients between the 1L-WS$_2$ flake and the SiO$_2$/Si substrate, we attribute this redshift primarily to thermal effects.

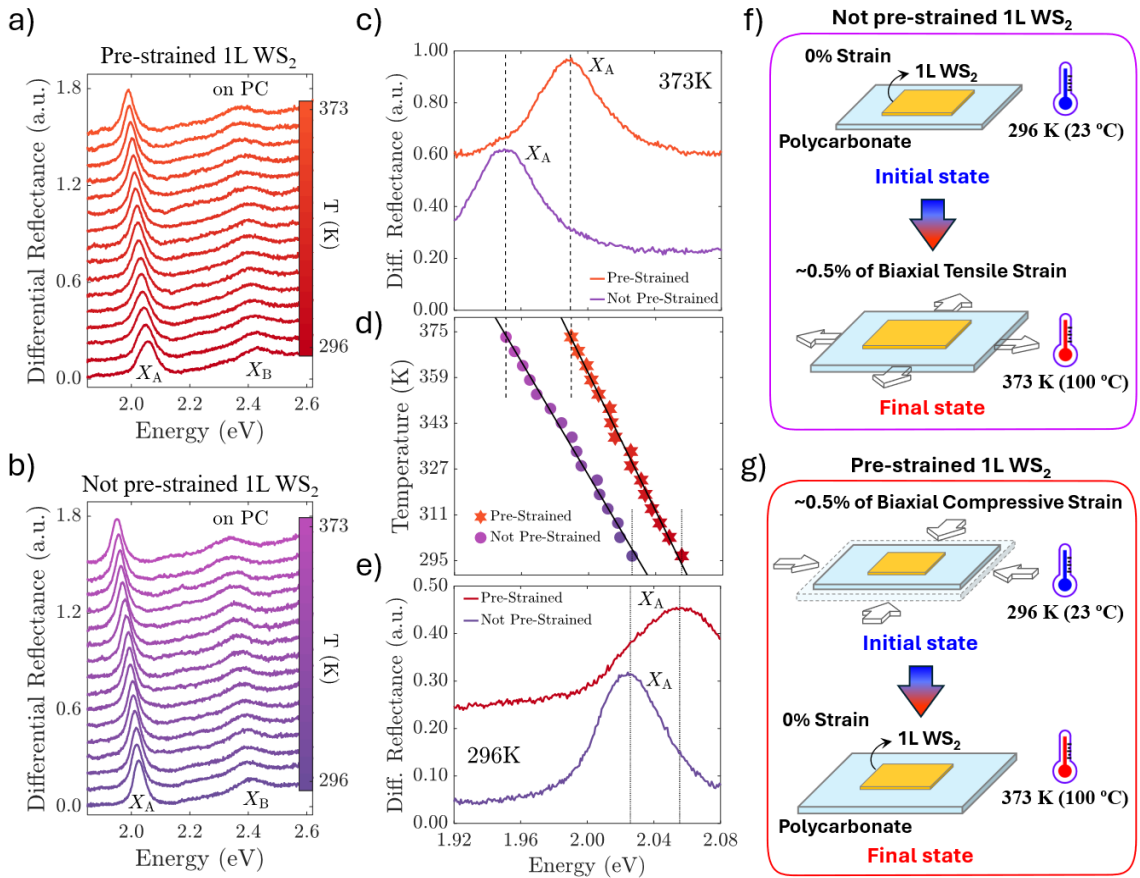

**Figure 2**. Differential reflectance as a function of temperature for single-layer WS$_2$ deposited onto a PC substrate via (a) the hot-dry-transfer method-pre-strained sample and (b) a deterministic dry transfer method at 296K (23ºC, not pre-strained sample). Spectra are vertically offset for clarity. (c) Zoom-in of the $X_A$ exciton resonance energy region from panels (a) and (b) at 373K (100ºC). (d) Energy positions for resonance $X_A$ extracted from panels (a) for pre-strained 1L-WS$_2$ and (b) for not pre-strained 1L-WS$_2$ as a function of temperature. The black solid lines represent linear fits to the data points. (e) Zoom-in of the $X_A$ resonance energy region from panels (a) and (b) at 296 K (23 ºC) for both samples. Dashed lines in panels (c-e) mark the energy position of the $X_A$ resonance for both samples. (f,g) Schematic diagram illustrating the strain states of the (f) not pre-strained 1L-WS$_2$ and (g) pre-strained 1L-WS$_2$ samples at 296 K (23 ºC) and 373 K (100 ºC).

In the case of samples on PC, at 23 ºC (296 K), the $X_A$ exciton peak appears at an energy ~30 meV higher in the pre-strained 1L-WS$_2$ sample compared to the not pre-strained one (**Figure 2e**). As the temperature increases up to 100 ºC, exciton peaks redshift similarly in both samples, with total shifts of ~60-70 meV, and slopes of ~-0.8 and ~-1 meV/K for the pre-strained and not pretrained samples, respectively (**Figure 2d**). This redshift in PC samples is roughly twice as large as that observed in the SiO$_2$/Si case and can be attributed to the combined effects of thermal bandgap narrowing and induced tensile strain in the PC-supported samples. First, the thermal bandgap narrowing is approximately linear in this temperature range, and accounts for a shift of -0.4 meV/K, as determined from the control 1L-WS$_2$ sample on SiO$_2$/Si (see **Supporting Information, Figure S4**). Second, the thermal expansion of the PC substrate induces tensile strain in both pre-strained and not pre-strained 1L-WS$_2$ samples. As illustrated in **Figure 2f**, the not pre-strained 1L-WS$_2$ sample evolves from an initially unstrained state to a tensile strain of up to ~0.5% upon heating from room temperature to 100 ºC. In contrast, in the pre-strained sample (**Figure 2g**), the substrate expansion compensates the ~-0.5% biaxial compressive pre-strain present at RT, leading to a nearly unstrained configuration at 100 °C. While the initial and final strain configurations differ for the two PC samples, the net effect of substrate expansion leads to comparable excitonic redshifts in both cases. The slightly smaller energy-temperature slope observed in the pre-strained sample may arise from imperfections at the substrate-sample interface introduced during the pre-straining process, resulting in less efficient strain transfer, or from sample to sample variations in gauge factor[10]. Finally, upon cooling back to RT, both samples return to their initial states, and the reflectance spectra resemble those shown in **Figure 2e**.

**Accumulated compressive strain upon cooling down pre-strained samples to 5K**

Next, we further cooled a pair of 1L-WS$_2$ samples on PC, one pre-strained and one not pre-strained, from room temperature (296 K) down to cryogenic temperatures (5 K), monitoring the evolution of their excitonic resonances via micro-reflectance spectroscopy (**Figure 3a-c**). Additionally, and to isolate thermal effects, we also tracked the temperature evolution of the $X_A$ and $X_B$ resonance energies for a control 1L-WS$_2$ sample deposited on SiO$_2$/Si (see **Supporting Information, Figures S5** and **S6**). To quantify both thermal and strain contributions, the energy positions of the excitonic resonances were extracted from differential reflectance spectra for all three samples (see the **Materials and Methods** section and **Supporting Information, Section S6** for details). The temperature evolution of the energy peak positions for resonances (bottom) $X_A$ and (top) $X_B$ in the 5-300 K range is compared in **Figure 3d** for the pre-strained and not pre-strained 1L-WS$_2$ samples on PC, along with the control sample deposited on SiO$_2$/Si.

On SiO$_2$/Si substrates, the exciton energies of 1L-WS$_2$ exhibit a blueshift upon cooling, consistent with the expected bandgap widening driven by reduced electron–phonon interactions and changes in atomic bond lengths at low temperatures[57,58]. In comparison, the not pre-strained 1L-WS$_2$ sample on PC shows a significantly larger blueshift of the excitonic peaks, reflecting the combined effects of intrinsic thermal bandgap evolution and compressive strain induced by PC substrate contraction upon cooling[10]. The impact of changes in the substrate dielectric screening can be disregarded since the dielectric constant of PC changes by only ~2% over the full range of temperatures studied[10,59].

In the pre-strained sample, an initial blueshift of ~30 meV in both the $X_A$ and $X_B$ excitons is already present at room temperature relative to the not pre-strained samples (**Figure 3d**). This energy difference is maintained throughout the cooling process, with a slightly larger blue-shift observed at low temperatures in the pre-strained 1L-WS$_2$ compared to the not pre-strained case (see **Figure 3c,d**). At the base temperature (5K), the total shift in exciton energy for the pre-strained sample reaches ~200 meV with respect to the SiO$_2$/Si control sample, due to the accumulated induced strain from the total substrate contraction. This enhanced strain arises from the combined contributions of the initial pre-strain induced during the hot-transfer process and the subsequent additional strain induced by the thermal contraction of the PC substrate during cooling from room to cryogenic temperatures. While the not pre-strained sample is solely subjected to the compression of the PC substrate from 296 K to 5 K, estimated at ~1.2%, the pre-strained sample is subjected to an effective total substrate compression of ~1.7%, over the full temperature range, from the hot transfer (373 K) down to the base temperature (5 K).

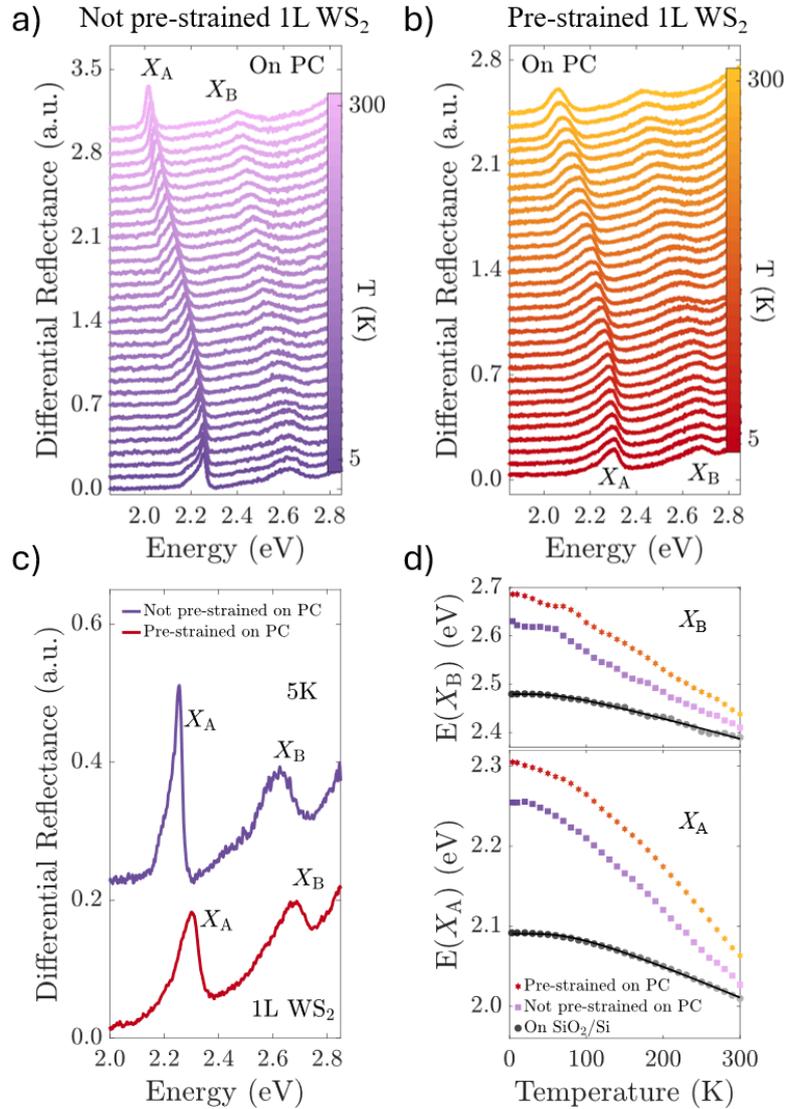

**Figure 3**. Differential reflectance as a function of temperature for (a) not pre-strained and (b) pre-strained single-layer WS$_2$ deposited on a polycarbonate (PC) substrate. Data for each sample are labelled in the top row of the figure. Spectra are vertically offset for clarity. (c) Comparison of differential reflectance spectra acquired at base temperature (5 K) for both samples. (d) Extracted energy peak positions for resonances $X_A$ (bottom) and $X_B$ (top), for 1L-WS$_2$ deposited on SiO$_2$/Si substrate (control sample) from **Supporting Information, Figure S6**, and from panel (a) for the not pre-strained 1L-WS$_2$, and panel (b) for the pre-strained 1L-WS$_2$ deposited on PC. In the case of data from 1L-WS$_2$ on SiO$_2$/Si substrate black solid lines represent fits to the temperature-dependent bandgap model from in Ref. [57].

The thermal evolution of the exciton peaks for the 1L-WS$_2$ deposited on SiO$_2$/Si was fitted to the O'Donell's model, which phenomenologically describes the temperature dependence of the bandgap energy for semiconductors[57,58] (see **Figure 3d and Supporting Information, Section S7**). We assume that data for 1L-WS$_2$ on SiO$_2$/Si comprises all purely temperature-related effects and can serve as a reference to isolate the effects arising from the induced biaxial compressive strain from purely thermal effects. To this end, we subtracted the fitted exciton energies in 1L-WS$_2$ on SiO$_2$/Si ($E_X^{SiO2/Si}$) from the exciton energies extracted for the pre-strained and not pre-strained 1L-WS$_2$ samples on PC ($E_X^{PC}$), yielding the strain-induced energy shifts $\Delta E = E_X^{PC} - E_X^{SiO2/Si}$. These values are presented as a function of substrate deformation for the $X_A$ exciton in the not pre-strained sample in **Figure S7** and for the pre-strained 1L-WS$_2$ in **Figure 4a** (in **Figure S8** for $X_B$). Although a small offset could arise due to the slightly different dielectric environments of SiO$_2$/Si and PC substrates at room temperature, this is expected to remain constant with temperature, as the dielectric constants of both materials do not vary significantly over the temperature range. We subtracted this constant offset to focus exclusively on the strain-induced component of ΔE.

The pre-strained sample exhibits a significantly larger ΔE than the control, resulting from the cumulative effect of up to ~0.5% compressive strain introduced during cooling from 373 to 296 K after hot-transfer, and up to ~1.2% additional compressive strain accumulated upon cooling from 296 K to 5 K. A total strain-induced ΔE ~ 200 meV was observed for the $X_A$ exciton, surpassing by ~50 meV the energy shift measured for not pre-strained 1L-WS$_2$ samples. This value represents the largest exciton energy shift reported to date for biaxial compressive strain in 1L-WS$_2$.

It is important to note that the deformation of the substrate provides an upper bound for the total strain transferred to the sample. Previous works estimate strain transfer efficiencies of ~ 80% from polycarbonate substrates to single-layer TMDCs[7,10]. The efficiency of the strain transfer can be quantified by a gauge factor, defined as the exciton energy shift per unit of substrate deformation. The ΔE-deformation relationship is typically assumed to be approximately linear according to theoretical expectations[60,61], and thus the gauge factor is usually defined as the slope of this linear dependence. However, the evolution of ΔE for $X_A$ exciton with substrate deformation in **Figure 4a** reveals a non-linear behaviour. This non-linearity may arise from a slightly non-linear dependence of exciton energy shift with strain, but also from strain transfer from substrate to flake being temperature dependent.

To quantify the temperature dependence of strain transfer efficiency, we calculated gauge factors by fitting the ΔE–deformation data with a polynomial function and evaluating its derivative. The resulting gauge factor values are shown as a function of temperature for $X_A$ and $X_B$ excitons in **Figure 4b**. Since gauge factors provide an experimental proxy for strain transfer efficiency, our results therefore indicate that strain transfer is less effective at high temperatures, resulting in moderate gauge factors. In contrast, at low temperatures, strain transfer improves significantly, with the gauge factor saturating near −150 meV/%. **Table 1** shows a comparison of the gauge factors obtained at 5 K with values predicted by theory, which can be considered the upper limit for ideal strain transfer. The fact that the gauge factor for pre-strained monolayer WS$_2$ on polycarbonate approaches the theoretical values at cryogenic temperatures can be interpret as evidence of near-perfect strain transfer.

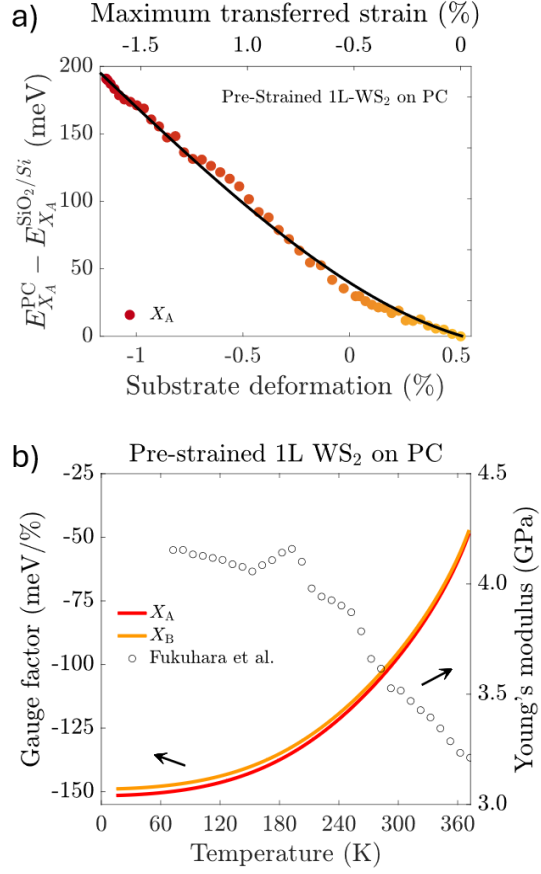

**Figure 4**. (a) Energy difference ΔE(X) = $E_X^{PC}$ – $E_X^{SiO2/Si}$ between the $X_A$ exciton peak positions in 1L-WS$_2$ transferred onto polycarbonate (PC) using a hot transfer method designed to induce pre-strain, and onto SiO$_2$/Si using a conventional dry transfer. ΔE(X) is plotted as a function of PC compression (ΔL/L, as estimated in **Figure S1**) and the corresponding maximum strain transferred to the flake. The black solid line corresponds to polynomial fit of the data. (b) Gauge factor for $X_A$ exciton, calculated as the derivative of the fit in panel (a). The Young's modulus of polycarbonate as a function of temperature is plotted for comparison (data is extracted from ref. [62]).

Two main parameters govern strain transfer from thermal mismatch with substrates: the thermal expansion coefficient and the Young's modulus. Although thermal expansion coefficients vary with temperature, this effect is already accounted for in the substrate deformation characterization as a function of temperature. Therefore, the observed temperature dependence of gauge factors is likely related to temperature variations in the Youngs modulus of polycarbonate. Indeed, while the reported values of Young's modulus for polycarbonate differ across studies[62–64], they consistently report a significant increase of this magnitude with decreasing temperature. **Figure 4b** overlays the Young's modulus of PC, compiled from Refs. [63,64], with the temperature-dependent gauge factors, revealing a striking correlation. At higher temperatures the Young´s modulus is lower, leading to reduced strain transfer efficiency. As the substrate stiffens with cooling, strain transfer improves, and gauge factors accordingly rise.

**Table 1**. Strain Gauge Factors Obtained from derivative of the polynomial fit of Exciton Energy shift (ΔE) versus Substrate Deformation in **Figure 4a** and **Figure S8**. Error is estimated from 95% confidence bounds.

| Gauge factor (meV/%) | $X_A$ | $X_B$ |
|---|---|---|
| This work | -148 ± 3 | -144 ± 3 |
| Theory | -144[61]<br>-151[7] | -123.8[61]<br>-130[7] |
| Henríquez-Guerra et al., ACS Appl. Mater. Interfaces, 2023[10] | -129 ± 3 | -112 ± 10 |
| Frisenda et al., npj 2D Mater. Appl., 2017[7] | -94[7]<br>-60* | -- |

*Gauge factor value after subtraction of temperature effects on the bandgap energy estimated in this work.

In summary, our results demonstrate that the efficiency of strain transfer from the substrate to 1L-WS$_2$ is predominantly governed by temperature-dependent variations in the polymer's Young's modulus. To further investigate this, we conducted a comparative study using polyamide 12 (PA-12)—a polymer with a higher thermal expansion coefficient but lower Young's modulus than polycarbonate (**see Supporting Information, Sections S10 and S11**). Despite undergoing greater thermal contraction (~1.9% vs. ~1.2%), 1L-WS$_2$ on PA-12 exhibited significantly smaller excitonic blue shifts during cooldown (ΔE ~90 meV), indicating less efficient strain transfer (see **Figures S11 and S12**). These findings reveal that a high thermal expansion coefficient alone is insufficient to achieve strong strain transfer; a sufficiently high Young's modulus is also essential. This dual requirement is consistent with previous studies[7,20] and highlights the critical role of mechanical properties—particularly the temperature dependence of Young's modulus—in strain transfer from thermal deformation of polymeric substrates.

CONCLUSIONS

In conclusion, we implemented a hot-dry transfer method onto thermally expanded polymer substrates, providing a simple and reversible strategy to fabricate samples with built-in compressive strain at room temperature. More importantly, this approach enables the induction of larger levels of uniform biaxial compressive strain in monolayer TMDCs upon cooling to cryogenic temperatures, expanding the experimentally accessible regime of compressive strain to up to ~1.7%. As a result of the accumulated strain from both the pre-straining step and further thermal contraction of the substrate, we observe excitonic blueshifts of up to 200 meV in 1L-WS$_2$—surpassing previous resports on compressive strain modulation in semiconductors. These findings establish hot-transfer pre-straining as a powerful strategy for inducing exceptionally large biaxial compressive strain in 2D materials at low temperatures. This opens new opportunities for strain-tuning low-temperature quantum phenomena such as excitonic behavior, superconductivity, and magnetism.

Our results also reveal a strong temperature dependence in strain transfer efficiency, with gauge factors increasing from moderate values at room temperature to 150 meV/% at 5 K, approaching the theoretical limit for ideal strain transfer. We attribute this enhancement to the increase in the Young's modulus of polycarbonate at low temperatures. By comparing results using different polymers, we confirmed that a high Young's modulus plays a more critical role than thermal expansion coefficient in enabling efficient strain transfer to 2D materials via thermal mismatch with the substrate, especially in the low temperature regime

MATERIALS AND METHODS

**Fabrication of single-layer WS$_2$ samples**

2H-WS$_2$ crystal (HQ Graphene) was mechanically exfoliated with Nitto 224 SPV tape onto transparent polydimethylsiloxane (Gel Film WF 4 x 6.0 mil from Gel-Pak®) substrate for inspection under an optical Motic BA310 metallurgical microscope. The exfoliation and inspection were carried out under ambient conditions. Differential reflectance measurements were performed at room temperature to identify single-layer samples. The selected samples were transferred using either a dry-transfer method[51] or a hot transfer method (at 373 K) onto a 250 µm thick polycarbonate (PC, by Modulor Gmbh, product number 0262951) film with dimensions of 6 mm × 6 mm. The PC substrate was selected for this hot transfer method due to its high glass transition temperature (423 K) and high melting point (568-588 K) compared to other polymers, such as polypropylene, polyethylene, or polyamide[52,53,65]. A control sample was simultaneously deposited onto a 6 mm × 6 mm SiO$_2$/Si substrate with a 50 nm oxide layer.

Polyamide 12 (PA-12) filaments from 3D printers were melted between two glass slides on a hot plate at 250-300 °C for 10 minutes. During the melting process, pressure was applied to the top of the glass slides using a 100 g mass to achieve a uniform surface. Once melted, the polymer was naturally cooled to room temperature while pressure was still applied. To separate the PA-12 films from the glass slides, they were cooled to 5 °C and then separated with the help of a blade. Films of approximately 1 mm thickness and dimensions of 6 mm × 6 mm were obtained.

**Room Temperature Raman Spectroscopy**

Raman spectra were obtained from single-layer WS$_2$ samples deposited on polycarbonate (PC) substrate. Measurements were performed under ambient conditions using a commercial Renishaw Raman spectrometer equipped with a 532 nm excitation laser. The laser beam was focused to a spot size of less than 1 µm in diameter, with an incident power of 25 µW. Each spectrum was acquired with an exposure time of 60 seconds and averaged over two accumulations.

**High Temperature Sample Stage**

A homemade stage was developed to apply biaxial tensile strain at high temperatures (>300 K). It consists of a brass plate with a heater (HT24S from Thorlabs) at the bottom, connected to a 20V power source, to ensure uniform heating over the entire surface of the plate. A thermistor (10K3A1I Series II from RS) embedded in the plate measures the temperature, and an Arduino-controlled solid-state relay and PID controller regulate the temperature to maintain a rate of approximately 1 K/min during the optical measurements. The brass plate is mounted on a stage with micromanipulators for precise movement during measurements.

**Thermal expansion/contraction of polymeric substrates**

The thermal expansion and contraction levels of PC and PA-12 substrates were extracted from the optical images of a lithographically micrometer scale pillar pattern at variable temperatures, following Refs.[7,10] (see also Supporting Information, Section S1).

**Variable Temperature Differential Reflectance Measurements**

Differential reflectance measurements were performed using a homemade microscope setup, based on ref.[66], and illuminated by a SOLIS-3C lamp (400–900 nm) from Thorlabs. A diaphragm focused the light to a spot size of approximately 3 μm on the sample, and the reflected light was collected by a tube lens and directed to a CCS200/M compact spectrometer (Thorlabs). Differential reflectance was calculated as DR = ($R_{sample}$ − $R_{substrate}$)/$R_{substrate}$. For the low temperature measurements, this setup was positioned at room temperature above a tabletop closed-cycle cryostat (AttoDry 800, Attocube GmbH), which controlled the sample temperature from 5 to 300 K. The cryostat, equipped with a piezoelectric controller, allowed for precise positioning of the sample. Reflectance spectroscopy measurements were recorded during both cooling and heating from 300 to 5 K (at approximately 1 K/min) under cryogenic vacuum (< 1 × $10^{-6}$ mbar) at the same sample and substrate locations.

**Determination of Energy Positions of Exciton Resonances**

The positions of the excitonic resonances are determined using the first and second derivative criteria as described by ref.[10]. Local maxima are identified where the first derivative of the reflectance spectrum equals zero, and the second derivative is negative. This method can be further validated by fitting the data to the Aspnes's equation[67]. The positions obtained from both methods show consistent agreement within a 5 meV margin of error.

ASSOCIATED CONTENT

**Supporting Information**

Supporting Information is provided below.

AUTHOR INFORMATION

**Corresponding Author**

Eudomar Henríquez-Guerra

eudomar.henriquez@bcmaterials.net

M. Reyes Calvo

reyes.calvo@bcmaterials.net

**Acknowledgements**

M.R.C. acknowledges funding from the Generalitat Valenciana Gent program (Cidegent2018/004), and from Ministry of Science and Innovation (Spain) through grants

PID2023-146354NB-C44, (funded by MICIU/AEI/10.13039/501100011033, and from EU FEDER). MRC and EHG from CNS2023-145151 (funded by MICIU/AEI/10.13039/501100011033 and from EU NextGenerationEU/PRTR). A. C-G. acknowledges funding by the Ministry of Science and Innovation (Spain) through the projects PDC2023-145920-I00 and PID2023-151946OB-I00 and the European Research Council through the ERC-2024-PoC StEnSo (grant agreement 101185235) and the ERC-2024-SyG SKIN2DTRONICS and the Severo Ochoa Centres of Excellence program through Grant CEX2024-001445-S. H.L. acknowledges support from China Scholarship Council (CSC) under grant 201907040070.

REFERENCES

(1) Qi, Y.; Sadi, M. A.; Hu, D.; Zheng, M.; Wu, Z.; Jiang, Y.; Chen, Y. P. Recent Progress in Strain Engineering on Van Der Waals 2D Materials: Tunable Electrical, Electrochemical, Magnetic, and Optical Properties. *Adv. Mater.* **2023**, *35* (12), 2205714. https://doi.org/10.1002/adma.202205714.

(2) Roldán, R.; Castellanos-Gomez, A.; Cappelluti, E.; Guinea, F. Strain Engineering in Semiconducting Two-Dimensional Crystals. *J. Phys. Condens. Matter* **2015**, *27* (31), 313201. https://doi.org/10.1088/0953-8984/27/31/313201.

(3) Pizzochero, M.; Yazyev, O. V. Inducing Magnetic Phase Transitions in Monolayer $CrI_3$ via Lattice Deformations. *J. Phys. Chem. C* **2020**, *124* (13), 7585–7590. https://doi.org/10.1021/acs.jpcc.0c01873.

(4) Wu, Z.; Yu, J.; Yuan, S. Strain-Tunable Magnetic and Electronic Properties of Monolayer $CrI_3$. *Phys. Chem. Chem. Phys.* **2019**, *21* (15), 7750–7755. https://doi.org/10.1039/C8CP07067A.

(5) Webster, L.; Yan, J.-A. Strain-Tunable Magnetic Anisotropy in Monolayer CrCl 3 , CrBr 3 , and CrI 3. *Phys. Rev. B* **2018**, *98* (14), 144411. https://doi.org/10.1103/PhysRevB.98.144411.

(6) Plechinger, G.; Castellanos-Gomez, A.; Buscema, M.; Van Der Zant, H. S. J.; Steele, G. A.; Kuc, A.; Heine, T.; Schüller, C.; Korn, T. Control of Biaxial Strain in Single-Layer Molybdenite Using Local Thermal Expansion of the Substrate. *2D Mater.* **2015**, *2* (1), 015006. https://doi.org/10.1088/2053-1583/2/1/015006.

(7) Frisenda, R.; Drüppel, M.; Schmidt, R.; Michaelis De Vasconcellos, S.; Perez De Lara, D.; Bratschitsch, R.; Rohlfing, M.; Castellanos-Gomez, A. Biaxial Strain Tuning of the Optical Properties of Single-Layer Transition Metal Dichalcogenides. *Npj 2D Mater. Appl.* **2017**, *1* (1), 10. https://doi.org/10.1038/s41699-017-0013-7.

(8) Gant, P.; Huang, P.; Pérez De Lara, D.; Guo, D.; Frisenda, R.; Castellanos-Gomez, A. A Strain Tunable Single-Layer MoS2 Photodetector. *Mater. Today* **2019**, *27*, 8–13. https://doi.org/10.1016/j.mattod.2019.04.019.

(9) Ryu, Y. K.; Carrascoso, F.; López-Nebreda, R.; Agraït, N.; Frisenda, R.; Castellanos-Gomez, A. Microheater Actuators as a Versatile Platform for Strain Engineering in 2D Materials. *Nano Lett.* **2020**, *20* (7), 5339–5345. https://doi.org/10.1021/acs.nanolett.0c01706.

(10) Henríquez-Guerra, E.; Li, H.; Pasqués-Gramage, P.; Gosálbez-Martínez, D.; D'Agosta, R.; Castellanos-Gomez, A.; Calvo, M. R. Large Biaxial Compressive Strain Tuning of Neutral and Charged Excitons in Single-Layer Transition Metal Dichalcogenides. *ACS Appl. Mater. Interfaces* **2023**, acsami.3c13281. https://doi.org/10.1021/acsami.3c13281.

(11) Henríquez-Guerra, E.; Almonte, L.; Li, H.; Elvira, D.; Calvo, M. R.; Castellanos-Gomez, A. Modulation of the Superconducting Phase Transition in Multilayer 2H-$NbSe_2$ Induced by Uniform Biaxial Compressive Strain. *Nano Lett.* **2024**, *24* (34), 10504–10509. https://doi.org/10.1021/acs.nanolett.4c02421.

(12) Henríquez-Guerra, E.; Ruiz, A. M.; Galbiati, M.; Cortés-Flores, Á.; Brown, D.; Zamora-Amo, E.; Almonte, L.; Shumilin, A.; Salvador-Sánchez, J.; Pérez-Rodríguez, A.; Orue, I.; Cantarero, A.; Castellanos-Gomez, A.; Mompeán, F.; Garcia-Hernandez, M.; Navarro-Moratalla, E.; Diez, E.; Amado, M.; Baldoví, J. J.; Calvo, M. R. Strain Engineering of Magnetoresistance and Magnetic Anisotropy in CrSBr. *Adv. Mater.* n/a (n/a), 2506695. https://doi.org/10.1002/adma.202506695.

(13) Liu, Z.; Amani, M.; Najmaei, S.; Xu, Q.; Zou, X.; Zhou, W.; Yu, T.; Qiu, C.; Birdwell, A. G.; Crowne, F. J.; Vajtai, R.; Yakobson, B. I.; Xia, Z.; Dubey, M.; Ajayan, P. M.; Lou, J. Strain and Structure Heterogeneity in MoS2 Atomic Layers Grown by Chemical Vapour Deposition. *Nat. Commun.* **2014**, *5* (1), 5246. https://doi.org/10.1038/ncomms6246.

(14) Kim, H.; Uddin, S. Z.; Lien, D.-H.; Yeh, M.; Azar, N. S.; Balendhran, S.; Kim, T.; Gupta, N.; Rho, Y.; Grigoropoulos, C. P.; Crozier, K. B.; Javey, A. Actively Variable-Spectrum Optoelectronics with Black Phosphorus. *Nature* **2021**, *596* (7871), 232–237. https://doi.org/10.1038/s41586-021-03701-1.

(15) Bertolazzi, S.; Gobbi, M.; Zhao, Y.; Backes, C.; Samorì, P. Molecular Chemistry Approaches for Tuning the Properties of Two-Dimensional Transition Metal Dichalcogenides. *Chem. Soc. Rev.* **2018**, *47* (17), 6845–6888. https://doi.org/10.1039/C8CS00169C.

(16) Raja, A.; Chaves, A.; Yu, J.; Arefe, G.; Hill, H. M.; Rigosi, A. F.; Berkelbach, T. C.; Nagler, P.; Schüller, C.; Korn, T.; Nuckolls, C.; Hone, J.; Brus, L. E.; Heinz, T. F.; Reichman, D. R.; Chernikov, A. Coulomb Engineering of the Bandgap and Excitons in Two-Dimensional Materials. *Nat. Commun.* **2017**, *8* (1), 15251. https://doi.org/10.1038/ncomms15251.

(17) Ramos, M.; Marques-Moros, F.; Esteras, D. L.; Mañas-Valero, S.; Henríquez-Guerra, E.; Gadea, M.; Baldoví, J. J.; Canet-Ferrer, J.; Coronado, E.; Calvo, M. R. Photoluminescence Enhancement by Band Alignment Engineering in $MoS_2$/$FePS_3$ van Der Waals Heterostructures. *ACS Appl. Mater. Interfaces* **2022**, *14* (29), 33482–33490. https://doi.org/10.1021/acsami.2c05464.

(18) Ramos, M.; Gadea, M.; Mañas-Valero, S.; Boix-Constant, C.; Henríquez-Guerra, E.; Díaz-García, M. A.; Coronado, E.; Calvo, M. R. Tunable, Multifunctional Opto-Electrical Response in Multilayer $FePS_3$/Single-Layer $MoS_2$ van Der Waals p–n Heterojunctions. *Nanoscale Adv.* **2024**, *6* (7), 1909–1916. https://doi.org/10.1039/D3NA01134H.

(19) Falin, A.; Holwill, M.; Lv, H.; Gan, W.; Cheng, J.; Zhang, R.; Qian, D.; Barnett, M. R.; Santos, E. J. G.; Novoselov, K. S.; Tao, T.; Wu, X.; Li, L. H. Mechanical Properties of Atomically Thin Tungsten Dichalcogenides: $WS_2$, $WSe_2$, and $WTe_2$. *ACS Nano* **2021**, *15* (2), 2600–2610. https://doi.org/10.1021/acsnano.0c07430.

(20) Liu, K.; Yan, Q.; Chen, M.; Fan, W.; Sun, Y.; Suh, J.; Fu, D.; Lee, S.; Zhou, J.; Tongay, S.; Ji, J.; Neaton, J. B.; Wu, J. Elastic Properties of Chemical-Vapor-Deposited Monolayer $MoS_2$, $WS_2$, and Their Bilayer Heterostructures. *Nano Lett.* **2014**, *14* (9), 5097–5103. https://doi.org/10.1021/nl501793a.

(21) Iguiñiz, N.; Frisenda, R.; Bratschitsch, R.; Castellanos-Gomez, A. Revisiting the Buckling Metrology Method to Determine the Young's Modulus of 2D Materials. *Adv. Mater.* **2019**, *31* (10), 1807150. https://doi.org/10.1002/adma.201807150.

(22) Bertolazzi, S.; Brivio, J.; Kis, A. Stretching and Breaking of Ultrathin $MoS_2$. *ACS Nano* **2011**, *5* (12), 9703–9709. https://doi.org/10.1021/nn203879f.

(23) Mueller, T.; Malic, E. Exciton Physics and Device Application of Two-Dimensional Transition Metal Dichalcogenide Semiconductors. *Npj 2D Mater. Appl.* **2018**, *2* (1), 29. https://doi.org/10.1038/s41699-018-0074-2.

(24) Mak, K. F.; Lee, C.; Hone, J.; Shan, J.; Heinz, T. F. Atomically Thin $MoS_2$: A New Direct-Gap Semiconductor. *Phys. Rev. Lett.* **2010**, *105* (13), 136805. https://doi.org/10.1103/PhysRevLett.105.136805.

(25) Chernikov, A.; Berkelbach, T. C.; Hill, H. M.; Rigosi, A.; Li, Y.; Aslan, B.; Reichman, D. R.; Hybertsen, M. S.; Heinz, T. F. Exciton Binding Energy and Nonhydrogenic Rydberg Series in

Monolayer WS 2. *Phys. Rev. Lett.* **2014**, *113* (7), 076802. https://doi.org/10.1103/PhysRevLett.113.076802.

(26) Zhu, B.; Chen, X.; Cui, X. Exciton Binding Energy of Monolayer WS2. *Sci. Rep.* **2015**, *5* (1), 9218. https://doi.org/10.1038/srep09218.

(27) Plechinger, G.; Nagler, P.; Kraus, J.; Paradiso, N.; Strunk, C.; Schüller, C.; Korn, T. Identification of Excitons, Trions and Biexcitons in Single-Layer WS$_2$: Identification of Excitons, Trions and Biexcitons in Single-Layer WS$_2$. *Phys. Status Solidi RRL - Rapid Res. Lett.* **2015**, *9* (8), 457–461. https://doi.org/10.1002/pssr.201510224.

(28) Zheng, W.; Jiang, Y.; Hu, X.; Li, H.; Zeng, Z.; Wang, X.; Pan, A. Light Emission Properties of 2D Transition Metal Dichalcogenides: Fundamentals and Applications. *Adv. Opt. Mater.* **2018**, *6* (21), 1800420. https://doi.org/10.1002/adom.201800420.

(29) Gao, L. Flexible Device Applications of 2D Semiconductors. *Small* **2017**, *13* (35), 1603994. https://doi.org/10.1002/smll.201603994.

(30) Zheng, L.; Wang, X.; Jiang, H.; Xu, M.; Huang, W.; Liu, Z. Recent Progress of Flexible Electronics by 2D Transition Metal Dichalcogenides. *Nano Res.* **2022**, *15* (3), 2413–2432. https://doi.org/10.1007/s12274-021-3779-z.

(31) Akinwande, D.; Petrone, N.; Hone, J. Two-Dimensional Flexible Nanoelectronics. *Nat. Commun.* **2014**, *5* (1), 5678. https://doi.org/10.1038/ncomms6678.

(32) Katiyar, A. K.; Ahn, J. Strain-Engineered 2D Materials: Challenges, Opportunities, and Future Perspectives. *Small Methods* **2024**, 2401404. https://doi.org/10.1002/smtd.202401404.

(33) Yu, X.; Peng, Z.; Xu, L.; Shi, W.; Li, Z.; Meng, X.; He, X.; Wang, Z.; Duan, S.; Tong, L.; Huang, X.; Miao, X.; Hu, W.; Ye, L. Manipulating 2D Materials through Strain Engineering. *Small* **2024**, 2402561. https://doi.org/10.1002/smll.202402561.

(34) Chaves, A.; Azadani, J. G.; Alsalman, H.; Da Costa, D. R.; Frisenda, R.; Chaves, A. J.; Song, S. H.; Kim, Y. D.; He, D.; Zhou, J.; Castellanos-Gomez, A.; Peeters, F. M.; Liu, Z.; Hinkle, C. L.; Oh, S.-H.; Ye, P. D.; Koester, S. J.; Lee, Y. H.; Avouris, Ph.; Wang, X.; Low, T. Bandgap Engineering of Two-Dimensional Semiconductor Materials. *Npj 2D Mater. Appl.* **2020**, *4* (1), 29. https://doi.org/10.1038/s41699-020-00162-4.

(35) Carrascoso, F.; Frisenda, R.; Castellanos-Gomez, A. Biaxial versus Uniaxial Strain Tuning of Single-Layer MoS2. *Nano Mater. Sci.* **2022**, *4* (1), 44–51. https://doi.org/10.1016/j.nanoms.2021.03.001.

(36) Carrascoso, F.; Li, H.; Obrero-Perez, J. M.; Aparicio, F. J.; Borras, A.; Island, J. O.; Barranco, A.; Castellanos-Gomez, A. Improved Strain Engineering of 2D Materials by Adamantane Plasma Polymer Encapsulation. *Npj 2D Mater. Appl.* **2023**, *7* (1), 24. https://doi.org/10.1038/s41699-023-00393-1.

(37) Michail, A.; Anestopoulos, D.; Delikoukos, N.; Parthenios, J.; Grammatikopoulos, S.; Tsirkas, S. A.; Lathiotakis, N. N.; Frank, O.; Filintoglou, K.; Papagelis, K. Biaxial Strain Engineering of CVD and Exfoliated Single- and Bi-Layer MoS$_2$ Crystals. *2D Mater.* **2021**, *8* (1), 015023. https://doi.org/10.1088/2053-1583/abc2de.

(38) Michail, A.; Anestopoulos, D.; Delikoukos, N.; Grammatikopoulos, S.; Tsirkas, S. A.; Lathiotakis, N. N.; Frank, O.; Filintoglou, K.; Parthenios, J.; Papagelis, K. Tuning the Photoluminescence and Raman Response of Single-Layer WS$_2$ Crystals Using Biaxial Strain. *J. Phys. Chem. C* **2023**, *127* (7), 3506–3515. https://doi.org/10.1021/acs.jpcc.2c06933.

(39) Li, H.; Carrascoso, F.; Borrás, A.; Moreno, G. P.; Aparicio, F. J.; Barranco, Á.; Gómez, A. C. Towards Efficient Strain Engineering of 2D Materials: A Four-Points Bending Approach for Compressive Strain. *Nano Res.* **2024**, *17* (6), 5317–5325. https://doi.org/10.1007/s12274-023-6402-7.

(40) Michail, A.; Yang, J. A.; Filintoglou, K.; Balakeras, N.; Nattoo, C. A.; Bailey, C. S.; Daus, A.; Parthenios, J.; Pop, E.; Papagelis, K. Biaxial Strain Transfer in Monolayer MoS$_2$ and WSe$_2$

Transistor Structures. *ACS Appl. Mater. Interfaces* **2024**, *16* (37), 49602–49611. https://doi.org/10.1021/acsami.4c07216.

(41) Roy, S.; Yang, X.; Gao, J. Biaxial Strain Tuned Upconversion Photoluminescence of Monolayer WS2. *Sci. Rep.* **2024**, *14* (1), 3860. https://doi.org/10.1038/s41598-024-54185-8.

(42) Shin, H.; Katiyar, A. K.; Hoang, A. T.; Yun, S. M.; Kim, B. J.; Lee, G.; Kim, Y.; Lee, J.; Kim, H.; Ahn, J.-H. Nonconventional Strain Engineering for Uniform Biaxial Tensile Strain in $MoS_2$ Thin Film Transistors. *ACS Nano* **2024**, *18* (5), 4414–4423. https://doi.org/10.1021/acsnano.3c10495.

(43) Harats, M. G.; Kirchhof, J. N.; Qiao, M.; Greben, K.; Bolotin, K. I. Dynamics and Efficient Conversion of Excitons to Trions in Non-Uniformly Strained Monolayer WS2. *Nat. Photonics* **2020**, *14* (5), 324–329. https://doi.org/10.1038/s41566-019-0581-5.

(44) Islam, M. A.; Nicholson, E.; Barri, N.; Onodera, M.; Starkov, D.; Serles, P.; He, S.; Kumral, B.; Zavabeti, A.; Shahsa, H.; Cui, T.; Wang, G.; Machida, T.; Singh, C. V.; Filleter, T. Strain Driven Electrical Bandgap Tuning of Atomically Thin $WSe_2$. *Adv. Electron. Mater.* **2024**, *10* (11), 2400225. https://doi.org/10.1002/aelm.202400225.

(45) Abramov, A. N.; Chestnov, I. Y.; Alimova, E. S.; Ivanova, T.; Mukhin, I. S.; Krizhanovskii, D. N.; Shelykh, I. A.; Iorsh, I. V.; Kravtsov, V. Photoluminescence Imaging of Single Photon Emitters within Nanoscale Strain Profiles in Monolayer WSe2. *Nat. Commun.* **2023**, *14* (1), 5737. https://doi.org/10.1038/s41467-023-41292-9.

(46) Lloyd, D.; Liu, X.; Christopher, J. W.; Cantley, L.; Wadehra, A.; Kim, B. L.; Goldberg, B. B.; Swan, A. K.; Bunch, J. S. Band Gap Engineering with Ultralarge Biaxial Strains in Suspended Monolayer $MoS_2$. *Nano Lett.* **2016**, *16* (9), 5836–5841. https://doi.org/10.1021/acs.nanolett.6b02615.

(47) Covre, F. S.; Faria, P. E.; Gordo, V. O.; De Brito, C. S.; Zhumagulov, Y. V.; Teodoro, M. D.; Couto, O. D. D.; Misoguti, L.; Pratavieira, S.; Andrade, M. B.; Christianen, P. C. M.; Fabian, J.; Withers, F.; Galvão Gobato, Y. Revealing the Impact of Strain in the Optical Properties of Bubbles in Monolayer $MoSe_2$. *Nanoscale* **2022**, *14* (15), 5758–5768. https://doi.org/10.1039/D2NR00315E.

(48) Iff, O.; Tedeschi, D.; Martín-Sánchez, J.; Moczała-Dusanowska, M.; Tongay, S.; Yumigeta, K.; Taboada-Gutiérrez, J.; Savaresi, M.; Rastelli, A.; Alonso-González, P.; Höfling, S.; Trotta, R.; Schneider, C. Strain-Tunable Single Photon Sources in $WSe_2$ Monolayers. *Nano Lett.* **2019**, *19* (10), 6931–6936. https://doi.org/10.1021/acs.nanolett.9b02221.

(49) Varghese, A.; Pandey, A. H.; Sharma, P.; Yin, Y.; Medhekar, N. V.; Lodha, S. Electrically Controlled High Sensitivity Strain Modulation in $MoS_2$ Field-Effect Transistors via a Piezoelectric Thin Film on Silicon Substrates. *Nano Lett.* **2024**, *24* (28), 8472–8480. https://doi.org/10.1021/acs.nanolett.4c00357.

(50) An, Z.; Soubelet, P.; Zhumagulov, Y.; Zopf, M.; Delhomme, A.; Qian, C.; Faria Junior, P. E.; Fabian, J.; Cao, X.; Yang, J.; Stier, A. V.; Ding, F.; Finley, J. J. Strain Control of Exciton and Trion Spin-Valley Dynamics in Monolayer Transition Metal Dichalcogenides. *Phys. Rev. B* **2023**, *108* (4), L041404. https://doi.org/10.1103/PhysRevB.108.L041404.

(51) Castellanos-Gomez, A.; Buscema, M.; Molenaar, R.; Singh, V.; Janssen, L.; Van Der Zant, H. S. J.; Steele, G. A. Deterministic Transfer of Two-Dimensional Materials by All-Dry Viscoelastic Stamping. *2D Mater.* **2014**, *1* (1), 011002. https://doi.org/10.1088/2053-1583/1/1/011002.

(52) Sehrawat, M.; Rani, M.; Bharadwaj, S.; Sharma, S.; Chauhan, G. S.; Dhakate, S. R.; Singh, B. P. Glass Transition Temperature Measurement of Polycarbonate Specimen by Dynamic Mechanical Analyser Towards the Development of Reference Material. *MAPAN* **2022**, *37* (3), 517–527. https://doi.org/10.1007/s12647-022-00572-3.

(53) Shamim, N.; Koh, Y. P.; Simon, S. L.; McKenna, G. B. Glass Transition Temperature of Thin Polycarbonate Films Measured by Flash Differential Scanning Calorimetry. *J. Polym. Sci. Part B Polym. Phys.* **2014**, *52* (22), 1462–1468. https://doi.org/10.1002/polb.23583.

(54) Huang, X.; Gao, Y.; Yang, T.; Ren, W.; Cheng, H.-M.; Lai, T. Quantitative Analysis of Temperature Dependence of Raman Shift of Monolayer WS2. *Sci. Rep.* **2016**, *6* (1), 32236. https://doi.org/10.1038/srep32236.

(55) Carrascoso, F.; Li, H.; Frisenda, R.; Castellanos-Gomez, A. Strain Engineering in Single-, Bi- and Tri-Layer MoS2, MoSe2, WS2 and WSe2. *Nano Res.* **2021**, *14* (6), 1698–1703. https://doi.org/10.1007/s12274-020-2918-2.

(56) Roy, R.; Agrawal, D. K.; McKinstry, H. A. Very Low Thermal Expansion Coefficient Materials. *Annu. Rev. Mater. Sci.* **1989**, *19* (1), 59–81. https://doi.org/10.1146/annurev.ms.19.080189.000423.

(57) O'Donnell, K. P.; Chen, X. Temperature Dependence of Semiconductor Band Gaps. *Appl. Phys. Lett.* **1991**, *58* (25), 2924–2926. https://doi.org/10.1063/1.104723.

(58) Tongay, S.; Zhou, J.; Ataca, C.; Lo, K.; Matthews, T. S.; Li, J.; Grossman, J. C.; Wu, J. Thermally Driven Crossover from Indirect toward Direct Bandgap in 2D Semiconductors: $MoSe_2$ versus $MoS_2$. *Nano Lett.* **2012**, *12* (11), 5576–5580. https://doi.org/10.1021/nl302584w.

(59) Anderson, W. E. *Measurements on Insulating Materials at Cryogenic Temperatures; Final Report*; National Bureau of Standards: Washington, DC, Electrosystems Div, 1980.

(60) Defo, R. K.; Fang, S.; Shirodkar, S. N.; Tritsaris, G. A.; Dimoulas, A.; Kaxiras, E. Strain Dependence of Band Gaps and Exciton Energies in Pure and Mixed Transition-Metal Dichalcogenides. *Phys. Rev. B* **2016**, *94* (15), 155310. https://doi.org/10.1103/PhysRevB.94.155310.

(61) Zollner, K.; Junior, P. E. F.; Fabian, J. Strain-Tunable Orbital, Spin-Orbit, and Optical Properties of Monolayer Transition-Metal Dichalcogenides. *Phys. Rev. B* **2019**, *100* (19), 195126. https://doi.org/10.1103/PhysRevB.100.195126.

(62) Mikio Fukuhara, M. F.; Asao Sampei, A. S. Low-Temperature Elastic Moduli and Dilational and Shear Internal Friction of Polycarbonate. *Jpn. J. Appl. Phys.* **1996**, *35* (5S), 3218. https://doi.org/10.1143/JJAP.35.3218.

(63) Hartwig, G. *Polymer Properties at Room and Cryogenic Temperatures*; Springer Science & Business Media, 1995.

(64) Mulliken, A. D.; Boyce, M. C. Mechanics of the Rate-Dependent Elastic–Plastic Deformation of Glassy Polymers from Low to High Strain Rates. *Int. J. Solids Struct.* **2006**, *43* (5), 1331–1356. https://doi.org/10.1016/j.ijsolstr.2005.04.016.

(65) Biron, M. *Thermoplastics and Thermoplastic Composites*; William Andrew, 2012.

(66) Frisenda, R.; Niu, Y.; Gant, P.; Molina-Mendoza, A. J.; Schmidt, R.; Bratschitsch, R.; Liu, J.; Fu, L.; Dumcenco, D.; Kis, A.; De Lara, D. P.; Castellanos-Gomez, A. Micro-Reflectance and Transmittance Spectroscopy: A Versatile and Powerful Tool to Characterize 2D Materials. *J. Phys. Appl. Phys.* **2017**, *50* (7), 074002. https://doi.org/10.1088/1361-6463/aa5256.

(67) Aspnes, D. E. Third-Derivative Modulation Spectroscopy with Low-Field Electroreflectance. *Surf. Sci.* **1973**, *37*, 418–442. https://doi.org/10.1016/0039-6028(73)90337-3.